\documentclass[12pt,aps,prd,showpacs]{revtex4}

\usepackage{epsfig}
\usepackage{graphicx}
\usepackage{bbm}
\usepackage{amsmath}
\usepackage{amsfonts}
\usepackage{amssymb}
\usepackage{slashed}
\usepackage{feynmf}

\newcommand{\nc}{\newcommand}
\nc{\be}{\begin{equation}}
\nc{\ee}{\end{equation}}
\nc{\bea}{\begin{eqnarray}}
\nc{\eea}{\end{eqnarray}}
\nc{\nn}{\nonumber}
\nc{\acom}[2]{ \left\{ #1,#2 \right\} }
\nc{\com}[2]{ \left[ #1,#2 \right] }
\nc{\dd}{^\dagger}
\nc{\ddp}{^{\dagger\prime}}
\nc{\ddpp}{^{\dagger\prime\prime}}
\nc{\pp}{^{\prime\prime}}
\nc{\ml}{M^\dagger M}
\nc{\mr}{MM\dd}
\nc{\explrp}{{\rm e}^{\frac{i}{2} \overset{\leftharpoonup}{\partial_z}
  \overset{\rightharpoonup}{\partial_{k_z}} }}
\nc{\exprlp}{{\rm e}^{\frac{i}{2} \overset{\leftharpoonup}{\partial_{k_z}}
  \overset{\rightharpoonup}{\partial_z} }}
\nc{\explrm}{ e^{-\frac{i}{2} \overset{\leftharpoonup}{\partial_z}
  \overset{\rightharpoonup}{\partial_{k_z}} }}
\nc{\exprlm}{ e^{-\frac{i}{2} \overset{\leftharpoonup}{\partial_{k_z}}
  \overset{\rightharpoonup}{\partial_z} }}
\nc{\lp}{\left(}
\nc{\rp}{\right)}
\nc{\CP}{{\cal CP}}
\nc{\CPd}{({\cal CP})^\dagger}
\nc{\Q}{{\cal Q}}
\nc{\Qd}{({\cal Q})^\dagger}
\nc{\col}{{\cal C}}
\nc{\cm}{{{\cal M}^2}}
\nc{\cs}{{\cal S}}

\nc{\rmi}{\textrm{i}}
\nc{\rmd}{\textrm{d}}

\def\Slash#1{#1\kern-0.55em\raise.05ex\hbox{/}}
\def\slash#1{#1\kern-0.5em\raise.05ex\hbox{{$\scriptstyle /$}}}

\newcommand{\ie}{{\it i.e.}}

\newcommand{\eg}{{\it e.g.}}

\newcommand{\cf}{{\it cf.}}

\newcommand{\eq}{Eq.}

\newcommand{\fig}{Fig.}

\newcommand{\Ref}{Ref.}
\newcommand{\Refs}{Refs.}
\newcommand{\Sec}{Sec.}

\begin{document}

\title{The Effective Matter Potential for Highly Relativistic Neutrinos}

\author{Thomas Konstandin}\email{konstand@theophys.kth.se}
\affiliation{Division of Mathematical Physics, Department of Physics,
School of Engineering Sciences, Royal Institute of Technology (KTH) --
AlbaNova University Center, Roslagstullsbacken 11, 106 91 Stockholm,
Sweden}

\author{Tommy Ohlsson}\email{tommy@theophys.kth.se}
\affiliation{Division of Mathematical Physics, Department of Physics,
School of Engineering Sciences, Royal Institute of Technology (KTH) --
AlbaNova University Center, Roslagstullsbacken 11, 106 91 Stockholm,
Sweden}

\date{\today}

\begin{abstract}
We investigate matter effects on highly relativistic neutrinos. The
self-energy of neutrinos is determined in an electron or neutrino
background taking into account resonance and finite width effects of
the gauge bosons. We find minor changes compared to the formerly used
formula for the propagator function and large deviations of the
effective width from the decay width of the gauge bosons considering
higher moments of the electron or neutrino distribution function.
\end{abstract}


\maketitle

\section{Introduction}

In the seminal work~\cite{Wolfenstein:1977ue}, Wolfenstein explored
the possibility of a change in the dynamics of neutrinos based on
coherent forward-scattering with surrounding matter
particles. Especially interesting is the case of a flavor sensitive
charged-current background leading to a change in the neutrino
oscillation probabilities. The scattering amplitude was determined
using the effective four-fermion Fermi interaction. In the picture of
Wolfenstein, coherent forward-scattering leads to a refractive
coefficient that accordingly changes the propagation of neutrinos in a
medium similar to an effective potential of strength
\be
V = \sqrt{2} \, G_F \,( N_e - N_{\bar e} ),
\label{wolf_result}
\ee
where $G_F$ denotes the Fermi coupling constant and $N_e$ and $N_{\bar
e}$ denote the electron and positron densities of the considered
medium, respectively.

Of course, this picture is not valid for arbitrarily high neutrino
energies. At an energy scale $E \simeq m_W \approx 10^2 \, {\rm GeV}$,
where $m_W$ is the W-boson mass, we expect the effective Fermi
theory to break down and the dynamics of the W-bosons to become
important. Energies of several GeV can indeed play a role in neutrino
physics, if one, \eg, considers high energetic cosmic
neutrinos~\cite{Lunardini:2000fy} that could be accessible by
experiments. However, this estimate is premature in the sense that the
kinematics of the problem has been neglected. Arguing that the energy
should be of order $m_W$ in the center-of-mass frame, one would come
to the estimate (assuming the electrons to be at rest)
\be
s = 2E_\nu \, m_e + m_e^2 \simeq 2E_\nu \, m_e \approx m_W^2,
\ee
where $m_e$ is the electron mass, which leads to
\be 
E_\nu \approx \frac{m_W^2}{2m_e} \approx 10^7 \, \textrm{GeV},
\label{estimate2}
\ee
while for scattering with another neutrino (with mass of
  order 1~eV)
\be 
E_\nu \approx \frac{m_Z^2}{2m_\nu} \approx 10^{13} \, \textrm{GeV}.
\label{estimate2_b}
\ee
Indeed, taking higher-order corrections in the neutrino energy into
account, one finds to next-to-leading order for relativistic
neutrinos~\cite{Notzold:1987ik}
\be
V = \sqrt{2} \, G_F \, \left[ (N_e - N_{\bar e}) - \frac{8\,
E_\nu}{3m_W^2} \left(\left<E_e\right> N_e + \left<E_{\bar e}\right>
N_{\bar e} \right) \right] + {\cal
O}\left(\left<E_e^2\right>,\left<E_{\bar e}^2\right>\right),
\label{estimate3}
\ee
where $\left<E_e\right>$ and $\left<E_{\bar e}\right>$
denote the average energies of the
electrons and positrons in the plasma, respectively.

In this paper, we are interested in the limit of very high energetic
neutrinos in matter, where the expansion used in \eq~(\ref{estimate2})
breaks down. In principle, one expects a resonance effect coming from
the generation of real W-bosons by anti-neutrinos of center-of-mass
energy $s \sim m_W^2$. The shape of the resonance depends on the
finite decay width of the W-boson. This effect has been taken into
account in \Refs~\cite{Lunardini:2000sw,Lunardini:2000fy} by the
educated guess that the finite width can be represented by just
including an effective propagator of the W-boson into
\eq~(\ref{wolf_result}), leading to
\be
V = \sqrt{2} G_F \, \left[ N_e \, f_0 (-s_W) - N_{\bar e} f_0(s_W)
\right]
\label{sem_result}
\ee
with the propagator function given by
\be
f_0(s_W) = \frac{1-s_W}{(1-s_W)^2 + \gamma_W^2},
\label{Wp_func}
\ee
where $s_W \equiv 2E_\nu m_e/m_W^2$ and $\gamma_W \equiv \Gamma_W/m_W$
denotes the width of the W-boson.

The goal of this paper is to derive the propagator function from
thermal field theory~\cite{LeBellac, Kapusta}. In addition, we give
arguments how a resummation of the series in \eq~(\ref{estimate3})
changes our results. In the next section, we adopt the formalism of
statistical quantum field theory as it was done in
\Refs~\cite{Notzold:1987ik,Pal:1989xs} and we present analytic as well
as numerical results. Furthermore, we apply it to the cases with and
without finite decay width. 
In a thermal background, a new energy scale
enters (this scale is of order 1~keV if we consider neutrinos in
the Sun up to order 1~MeV for the cosmic neutrino background at
decoupling) that will change the influence of the finite decay width
of the gauge boson.  Finally, in \Sec~\ref{sec:S&C}, we summarize and
conclude our results.

\section{The Neutrino Self-Energy in a Statistical Background}

\begin{figure}[htbp]
\begin{center}
\includegraphics[width=6.0cm]{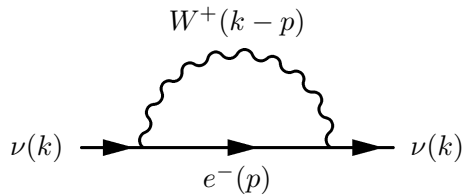}
\end{center}
\vskip -.8cm
\caption{%
\label{fig_fm}
The Feynman diagram for the charged-current contribution to the neutrino
self-energy.}
\end{figure}

In the following, we carry out the calculations for the neutrino
self-energy as was done in a temperature system in
\Refs~\cite{Weldon:1982bn,Petitgirard:1991mf,Quimbay:1995jn} and
applied to neutrinos in an electron background in
\Refs~\cite{Notzold:1987ik,Pal:1989xs}. In contrast to these works,
we expand in moments of the electron momentum $\left<p^n\right>$
instead of the electron energy $\left<E_e^n\right>$ what is better
suited for the investigation of the resonance regime. The
corresponding Feynman diagram is depicted in \fig~\ref{fig_fm}. The
self-energy contributions coming from other species such as neutrinos
or nuclear matter can be carried out along the same lines.
\vskip .5cm

In the real-time formalism, the electron propagator reads
\be
 S(p) = (\slashed{p} + m_e) \left[ 
\frac{1}{p^2 - m^2_e + \rmi\epsilon} + \rmi\Gamma(p) \right],
\ee
where $\Gamma$ describes the electron background and is given by
\be
\Gamma(p) = 2\pi\, \delta(p^2 -m_e^2) [ \theta(p^0) n_+(p^0) +
\theta(-p^0) n_-(-p^0) ]
\ee
and $n_\pm(p^0)$ denote the particle distribution functions of
electrons and positrons, respectively. The distribution functions are
usually parameterized by the temperature and chemical potentials
according to
\be
n_\pm(p_0) = \frac{1}{\textrm{e}^{(p_0-\mu_\pm)/T} + 1}.
\ee
Furthermore, in the unitary gauge, the W-boson propagator is given by
\be
\Delta_{\mu\nu}(q) = \frac{-1}{q^2 - m_W^2 + \rmi \epsilon}
\left(g_{\mu\nu} - \frac{q_\mu q_\nu}{m_W^2} \right).
\label{W_prop_1}
\ee

The first term in the electron propagator is the usual zero
temperature ($T=0$) part, while the second term gives the following
temperature contribution to the neutrino self-energy:
\bea
\Sigma_T(k) &=& -\frac{g^2}{2}\int \frac{\rmd^4p}{(2\pi)^3} \, \delta(p^2
-m_e^2) [\theta(p^0) n_+(p^0) + \theta(-p^0) n_-(-p^0) ] \nn\\
&& \quad \times
\gamma_\mu \, P_L \,
\Delta^{\mu\nu}(k-p) \, \slashed{p} \gamma_\nu \, P_L \,,
\label{eq_a}
\eea
where the weak interaction coupling constant $g$ is related to the
Fermi coupling constant $G_F$ as $g^2/(8 m_W^2) = G_F/\sqrt{2}$ and
$P_L \equiv (1 - \gamma_5)/2$ denotes the left-handed projection
operator.
Performing the evaluation of the Dirac algebra, this can be
transformed into~\cite{D'Olivo:1992vm}
\be
\Sigma_T(k) = -\frac{g^2}{2m_W^2} \left[ (2m_W^2 + m_e^2 - k^2)
\gamma_\mu {\cal P}^\mu + 2( k\cdot {\cal P} - m_e^2{\cal Q} )
\slashed{k}\right] P_L
\ee
with the definitions
\bea
{\cal Q} &=& \int \frac{\rmd^4p}{(2\pi)^3} [\theta(p^0) n_+(p^0) +
\theta(-p^0) n_-(-p^0) ] \, \frac{\delta(p^2 -m_e^2)}{(k-p)^2 - m^2_W +
\rmi\epsilon }, \\
{\cal P}^\mu &=& \int \frac{\rmd^4p}{(2\pi)^3} [\theta(p^0) n_+(p^0) +
\theta(-p^0) n_-(-p^0) ] \, \frac{p^\mu \delta(p^2 -m_e^2)}{(k-p)^2 -
m^2_W + \rmi\epsilon}.
\eea
The expression ${\cal P}^\mu$ can be split according to 
\be
{\cal P}^\mu = {\cal P}_u u^\mu + {\cal P}_k k^\mu,
\ee
where $u^\mu$ denotes the plasma four-vector that is $(1,0,0,0)$ in
the rest frame of the plasma.
The terms in the neutrino self-energy proportional to $\slashed{k}$
are related to the dispersion relation and the wave-function
normalization and they give only corrections of order $g^4$ to the
effective potential of the neutrinos~\cite{D'Olivo:1992vm}.
In what follows, we only analyze the resonance behavior of the
leading-order contribution to the neutrino self-energy
which translates into the following effective potential for highly
relativistic neutrinos~\cite{Mannheim:1987ef}
\be
V(k) = - g^2 \frac{2m_W^2 + m_e^2 - k^2}{2m_W^2}{\cal P}_u
\simeq - g^2 {\cal P}_u = - g^2 \frac{ k \cdot P}{k\cdot u}.
\label{simple}
\ee
For simplicity, we neglect the mass (thermal and non-thermal) of the
neutrino and we choose a frame such that $k=(E_\nu,0,0,E_\nu)$. Out
of the four integrations in \eq~(\ref{eq_a}), three can be performed
analytically, yielding
\bea
V(k) &=& \frac{g^2}4 \int_0^\infty \frac{\rmd p \, p^2}{2\pi^2
E_\nu E_e} \, n_+( E_e) \left[ 1 + \frac{m_W^2-m_e^2}{4p\,E_\nu}\ln
\left| \frac{m_W^2 - m_e^2 + 2E_\nu(E_e - p)} {m_W^2 - m_e^2 +
2E_\nu(E_e + p)} \right| \right] \nn \\
&&+\, \frac{g^2}4
\int_0^\infty \frac{\rmd p \, p^2}{2\pi^2 E_\nu E_e} \, n_-( E_e)
\left[ 1 + \frac{m_W^2-m_e^2}{4p\,E_\nu}\ln \left| \frac{m_W^2 - m_e^2
- 2E_\nu(E_e + p)} {m_W^2 - m_e^2 - 2E_\nu(E_e - p)} \right| \right]
\label{V_num}
\eea
with $E_e=\sqrt{m_e^2 + p^2}$. 

First, we will reproduce the result in \eq~(\ref{estimate3}). Thus, we
expand the logarithm in the loop-momentum $p$
\be
\ln
\left| \frac{m_W^2 - m_e^2 + 2E_\nu(E_e \mp p)} {m_W^2 - m_e^2 +
2E_\nu(E_e \pm p)} \right| \approx \mp \frac{4E_\nu p}{m_W^2 - m_e^2 -2E_\nu m_e} 
\ee
and obtain
\be
V(k) = \frac{g^2}4 \left( \frac{ N_e}{m_W^2 - m_e^2
+2E_\nu m_e} - \frac{ N_{\bar e}}{m_W^2 - m_e^2 -2E_\nu m_e} \right)
+ {\cal O} \left(\left<p \right>\right),
\label{res_1}
\ee
where we have defined
\be
N_e = \int_0^\infty \frac{\rmd p \, p^2}{2\pi^2} \, \, n_+( E_e),
\qquad N_{\bar e} = \int_0^\infty \frac{\rmd p \, p^2}{2\pi^2} \, \,
n_-( E_e)
\ee
as well as 
\be
\left<p^n\right>_e = \frac{1}{N_e}\int_0^\infty \frac{\rmd p \,
p^{n+2}}{2\pi^2} \, \, n_+( E_e), \qquad \left<p^n\right>_{\bar e} =
\frac{1}{N_{\bar e}} \int_0^\infty \frac{\rmd p \, p^{n+2}}{2\pi^2} \,
\, n_-( E_e).
\ee
Note that in the limit 
\be
\left<p^n\right> =0, \quad \left<E_e^n\right> =m_e^n,
\ee
this does not reproduce the result in \eq~(\ref{estimate3}), 
since there is an additional contribution
of order $\left<E_e^{-1}\right>$ that can be neglected for high
temperatures which was already pointed out in
\Ref~\cite{Notzold:1987ik}. Including this contribution (and
neglecting the positron background) yields
\be
{\cal P}_u = \frac{N_e}{m_W^2 - m_e^2} \left[ -\frac14 +
\frac{E_\nu}{6(m_W^2 - m_e^2)} \left( 4 \left<E_e\right> - m_e^2
\left<E_e^{-1}\right> \right) \right],
\ee
which agrees with our result in the considered limit.

\begin{figure}[htbp]
\begin{center}
\includegraphics[width=6.0in,clip]{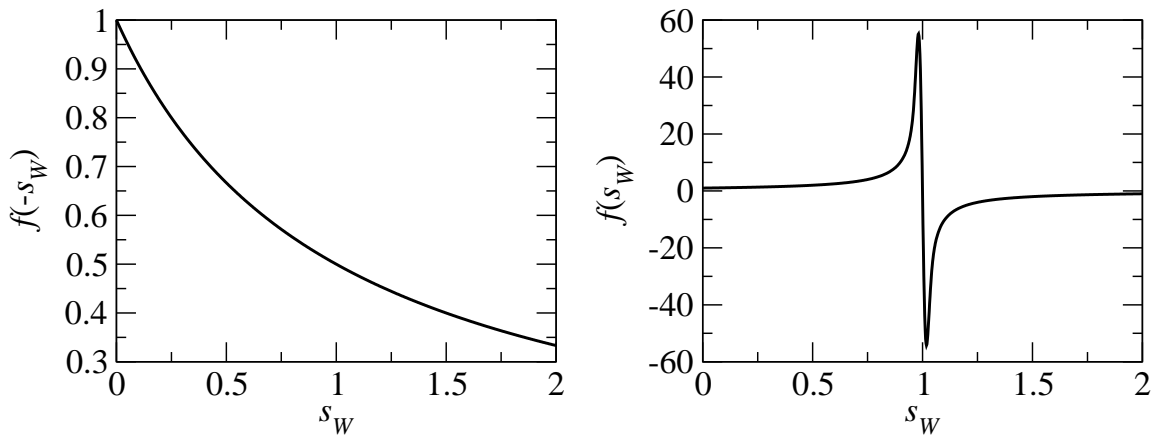}
\end{center}
\vskip -0.2in
\caption{%
\label{fig_num}
The propagator function $f(s_W)$ 
for neutrinos (left figure) and anti-neutrinos (right
figure) without finite decay width. The electron background
corresponds to the properties of the Sun.}
\end{figure}

So far, we have only expanded in the loop momentum $p$. The
convergence of this expansion depends on the quotient of higher
moments $\left< p^n \right>$ and the resonance condition appearing in
the denominators
\be
\frac{E^n_\nu \left< p^n \right>}{(m_W^2 - m_e^2 -2E_\nu m_e)^n} \ll 1. 
\ee
Hence, close to the resonance, our approximation breaks down and the
divergence in \eq~(\ref{res_1}) is removed due to the finite values of
the temperature $T$ and the chemical potential $\mu$ in the electron
distribution function. In \fig~\ref{fig_num}, the full momentum
dependence of the self-energy of neutrinos and anti-neutrinos is shown
for the electron distribution in the Sun.

The height of the resonance can be estimated as follows. Suppose that
the temperature $T$ of the distribution function is not much larger
than the chemical potential $\mu$ and notice that in this case the
higher-order moments behave as $\left< p^n \right> \sim \mu^n$. The
potential as given in \eq~(\ref{res_1}) then depends on the
center-of-mass energy $s$ and the combination $E_\nu \mu/m_W^2$. Being
close to the resonance and identifying this parameter with the width
as in \eq~(\ref{Wp_func}), we obtain the estimate
\be
\gamma_\mu^2 = \frac{E_\nu \mu}{m_W^2} \approx \frac{\mu}{2m_e}
\approx \left( \frac{1}{30} \right)^2.
\label{mu_gamma}
\ee
The numerical value is $\gamma_\mu \simeq 0.0107$, which is obtained by
fitting the result with a resonance shape of the form in
\eq~(\ref{Wp_func}). Hence, the estimate in \eq~(\ref{mu_gamma})
gives the right order of magnitude.

Second, we would like to examine the inclusion of a finite width of
the W-boson. The transverse and longitudinal parts of the propagator
in \eq~(\ref{W_prop_1}) then changes to~\cite{Calderon:2001qq}
\bea
\Delta^T_{\mu\nu}(q) &=& \frac{-1}{q^2 - m_W^2 + \rmi q^2 \gamma_W}
\left(g_{\mu\nu} - \frac{q_\mu q_\nu}{q^2} \right), \\
\Delta^L_{\mu\nu}(q) &=& \frac{1}{q^2 - m_W^2 + \rmi \epsilon}
\frac{q_\mu q_\nu}{q^2},
\label{W_prop_2}
\eea
where we have defined the W-boson width as $\gamma_W \equiv \Gamma_W /
m_W$. Using the same arguments that led us to \eq~(\ref{simple}), we
deduce that only the term proportional to the metric $g_{\mu\nu}$
contributes to leading order and that the longitudinal part of the
propagator can be neglected. Analogously to \eq~(\ref{V_num}),
we obtain
\bea
V(k) &=& \frac{g^2}4 \int_0^\infty \frac{\rmd p \, p^2}{2\pi^2
E_\nu E_e} \, n_+( E_e) \,\,\Re \left[ \frac{1}{z} + 
\frac{m_W^2}{4p\,E_\nu \, z^2}\ln
\left| \frac{m_W^2 + 2E_\nu(E_e - p) z } {m_W^2 +
2E_\nu(E_e + p) z } \right| \right] \nn \\
&&+\, \frac{g^2}4
\int_0^\infty \frac{\rmd p \, p^2}{2\pi^2 E_\nu E_e} \, n_-( E_e)
 \, \Re \left[ \frac{1}{z} + \frac{m_W^2}{4p\,E_\nu\, z^2}\ln \left| \frac{m_W^2 
- 2E_\nu(E_e + p)z} {m_W^2 - 2E_\nu(E_e - p)z} \right| \right],
\label{V_num2}
\eea
where we used the approximation $m_W^2 - m_e^2\approx m_W^2 $ and the
convention $z=1+ \rmi \gamma_W$ to shorten the expression.
Performing similar steps as before and
neglecting higher-order corrections in the temperature and the
chemical potential of the electrons, we find
\be
V(k) = \sqrt{2} \, G_F \, (N_e - N_{\bar e}) \, \Re \left[
\frac{m_W^2}{m_W^2 - E_\nu m_e \, z} \right],
\ee
or equivalently, the W-boson propagator function is given by 
\be
f (s_W) = \frac{1-s_W}{(1-s_W)^2 +s_W^2 \gamma_W^2}.
\label{Wp_func2}
\ee
in contrast to \eq~(\ref{Wp_func}). Note that, in the limit $s_W \to
0$, our result leads to the result of Wolfenstein, see
\eq~(\ref{wolf_result}), while \eq~(\ref{Wp_func}) gives a correction
by the factor $1/(1+\gamma_W^2)$. On the other hand, in the limit $s_W \to \infty$,
we obtain an additional factor $1/(1+\gamma_W^2)$ in the effective mass of the 
neutrinos.
This difference is due to the fact
that we used a $q^2$-dependent damping term in \eq~(\ref{W_prop_2}),
while a constant damping term can also be found in the literature.
However, one has to have in mind that for the Standard Model value
($\gamma_W \simeq 0.0266$) this correction is still out of the range
of experimental testability, since absorption processes become strong
close to the resonance~\cite{Lunardini:2000sw}. On the other hand, the
smallness of the width makes its influence compatible with the effect
coming from finite temperature and chemical potential, \cf
~\eq~(\ref{mu_gamma}).

\begin{figure}[htbp]
\begin{center}
\includegraphics[width=4.0in,clip]{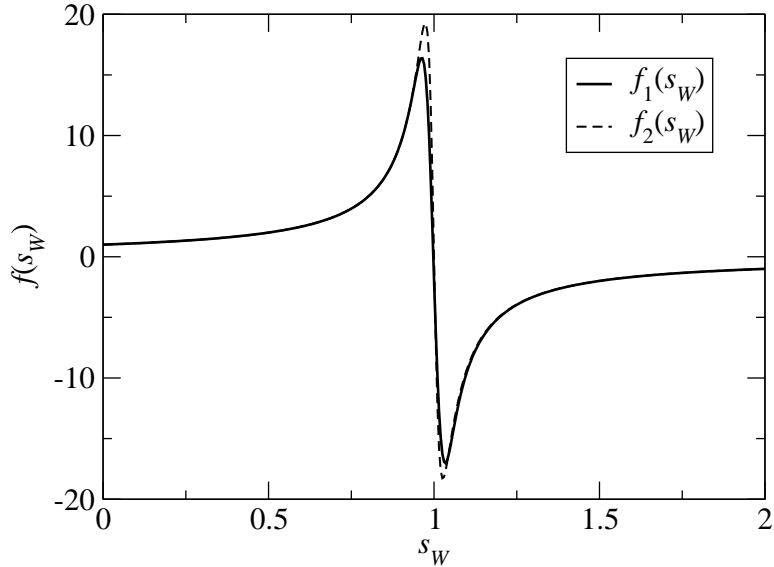}
\end{center}
\vskip -.2in
\caption{%
\label{fig_num2}
The propagator function $f(s_W)$ with finite W-boson decay width. 
The plot $f_1(s_W)$ shows the numerical result of \eq~(\ref{V_num2})
including finite temperature and chemical potential
effects, corresponding to $\gamma_{\textrm{eff}}\simeq0.0311$, while
the plot $f_2(s_W)$ shows the result of \eq~(\ref{Wp_func2}). In both
cases, the electron background meets the properties of the Sun.}
\end{figure}

In \fig~\ref{fig_num2}, we have plotted the result of
\eq~(\ref{Wp_func2}) and the full numerical result including the
effects of finite temperature and chemical potential that can be obtained 
analogously to \eq~(\ref{V_num}). The numerical
result corresponds to an effective width of
$\gamma_{\textrm{eff}}\simeq 0.0311$.

\begin{figure}[htbp]
\begin{center}
\includegraphics[width=4.0in,clip]{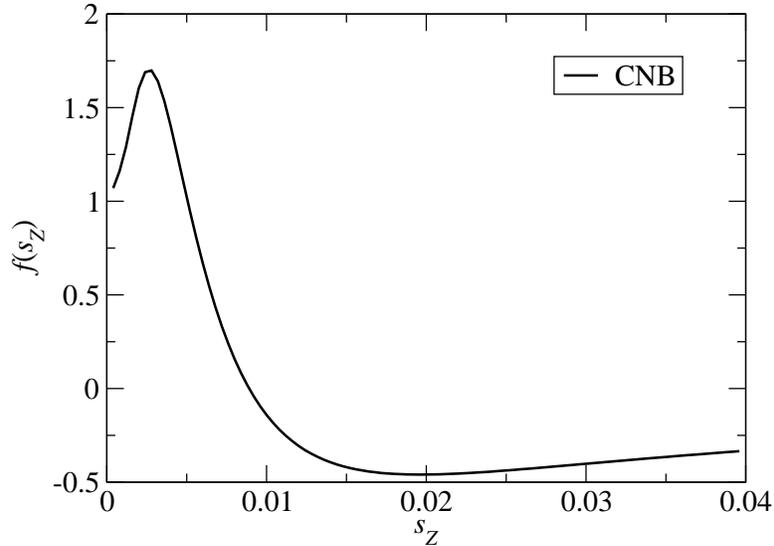}
\end{center}
\vskip -.2in
\caption{%
\label{fig_num3}
The propagator function $f(s_Z)$ with finite Z-boson decay width. The
plot $f(s_Z)$ shows the numerical result of \eq~(\ref{V_num2})
including finite temperature effects ($T = 10 \textrm{ eV}$) of the
cosmic neutrino background (CNB) of a neutrino of mass $m_\nu = 1\,$ eV
 without chemical potentials $\mu_\pm=0$. }
\end{figure}

Thus, we arrive at the conclusion that in the Sun for the
charged-current contributions the effect coming from the finite width
of the W-boson is larger than the temperature effect, \ie,
$\gamma_{\textrm{eff}} \approx \gamma_W \gtrsim \gamma_\mu$. This
picture can change in different contexts. For example, in supernovae,
the rather large chemical potential of the electrons leads to
$\gamma_\mu \approx 1$, and hence, the temperature effects dominate.
Similarly, this effect is important considering the scattering with
the neutrino background. Due to the small neutrino masses, the resonance
appears for very high neutrino energies. Thus, using
\eq~(\ref{mu_gamma}), we obtain
\be
\gamma_\mu^2 = \frac{E_\nu \mu}{m_W^2} \approx \frac{\mu}{2m_\nu} 
%\gtrsim 1
\ee
and the finite width from the Z-boson decay channels plays only a
marginal role as long as the density of the neutrino background is
larger than the mass scale of the neutrinos. This expectation is met
in the numerical results of \fig~\ref{fig_num3}, where the resonance
shape of the effective potential due to a neutrino background of
temperature $T = 10 \textrm{ eV}$ is plotted for a neutrino of
mass $m_\nu = 1\,$ eV and without chemical potentials $\mu_\pm=0$.
The resonance effect of the finite Z-width is almost completely
washed-out, while the maximum of the resonance is shifted to smaller
center-of-mass momentum.
Since neutrinos are relativistic most of the
time during the cosmological evolution (and one flavor could be
relativistic even today), we expect the change of the effective width
to be relevant for neutrinos from cosmic sources close to the resonance.

\section{Summary and Conclusion} 
\label{sec:S&C}

We have analyzed the rare scenarios, where the neutrino energy is
large enough such that the energy and momentum dependence of matter
effects could be relevant. This is, \eg, the case for neutrinos from
cosmic sources that has been analyzed in \Ref~\cite{Lunardini:2000fy}.
In this limit, we have derived the shape of the resonance from
thermal field theory and found a small deviation from the formerly
proposed propagator function. In addition, we have examined the change
in the effective width due to the finite temperature and chemical
potential and found that the effective width will be much larger than
the decay width of the W-boson in supernovae or the Z-boson width in
the case of scattering with the relic neutrino background. Finally, we
note that the Mikheyev--Smirnov--Wolfenstein
(MSW)~\cite{Wolfenstein:1977ue,Mikheev:1986gs,Mikheev:1986wj}
resonance condition for effective mixing of neutrinos in matter will,
in principle, be modified due to the temperature effects, since these
effects can be interpreted as flavor conserving non-standard
Hamiltonian effects~\cite{Blennow:2005qj}. However, the change will be
completely negligible for solar and supernova neutrinos.

\section*{Acknowledgments}

We would like to thank Mattias Blennow and Tomas H{\"a}llgren for
useful discussions. This work was supported by the Royal Swedish
Academy of Sciences (KVA), Swedish Research Council
(Vetenskapsr{\aa}det), Contract No.~621-2001-1611, 621-2002-3577, and
the G{\"o}ran Gustafsson Foundation (G{\"o}ran Gustafssons Stiftelse).

\end{document}